\documentclass[reprint,superscriptaddress,aps,prb,twocolumn]{revtex4-1}
\usepackage{setspace}
\usepackage{amsmath}
\usepackage{graphicx}
\usepackage{amsfonts}
\usepackage{amssymb}
\usepackage{epstopdf} 
\usepackage{xcolor}
\usepackage{verbatim}
\usepackage{soul}
\usepackage{comment}
\usepackage{placeins}
\usepackage{textcomp}

\usepackage{pdfpages} 
\usepackage{pgffor} 

\makeatletter
\AtBeginDocument{\let\LS@rot\@undefined}
\makeatother

\def\supplementfilename{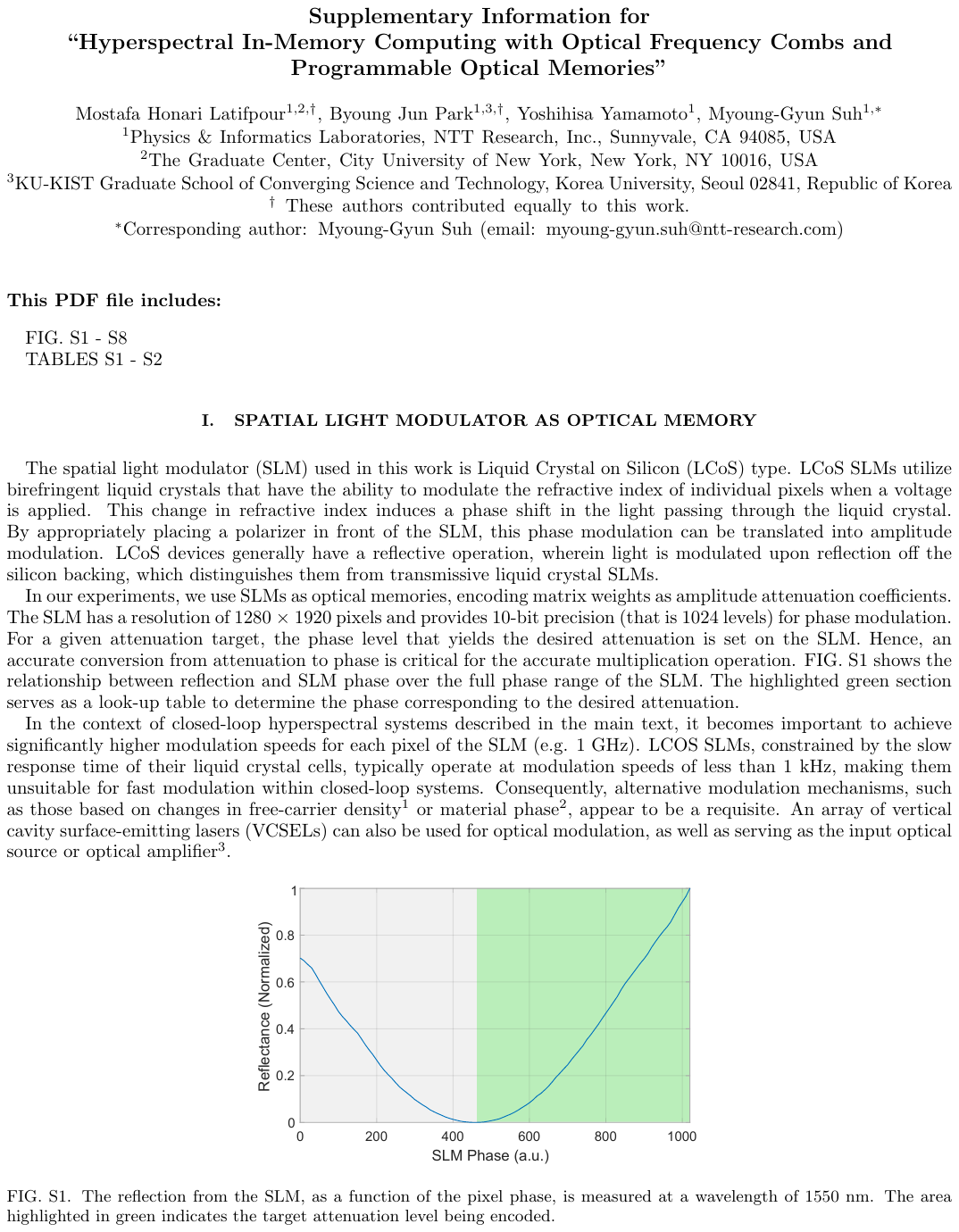}

\pdfximage{\supplementfilename}
\def\numbersupplementpages{\the\pdflastximagepages}

\newif\ifarXiv
\arXivtrue 

\usepackage[pagewise]{lineno}

\DeclareGraphicsExtensions{.pdf,.eps,.png,.jpg,.mps} 

\usepackage{float}

\begin{document}
\title{Hyperspectral In-Memory Computing\\with Optical Frequency Combs and Programmable Optical Memories}

\author{ Mostafa Honari Latifpour$^{1,2,\dagger}$, Byoung Jun Park$^{1,3,\dagger}$, Yoshihisa Yamamoto$^{1}$, Myoung-Gyun Suh$^{1,*}$\\
$^1$Physics $\&$ Informatics Laboratories, NTT Research, Inc., Sunnyvale, CA 94085, USA \\
$^2$The Graduate Center, City University of New York, New York, NY 10016, USA \\
$^3$KU-KIST Graduate School of Converging Science and Technology, Korea University, Seoul 02841, Republic of Korea\\
$^\dagger$ These authors contributed equally to this work.\\
$^*$Corresponding author: Myoung-Gyun Suh (email: myoung-gyun.suh@ntt-research.com)
}

\maketitle

\noindent{\textbf{ 
The rapid advancements in machine learning across numerous industries have amplified the demand for extensive matrix-vector multiplication operations\cite{krizhevsky2012imagenet,vaswani2017attention,lecun2015deep}, thereby challenging the capacities of traditional von Neumann computing architectures. To address this, researchers are currently exploring alternatives such as in-memory computing systems to develop faster and more energy-efficient hardware\cite{tsai2018recent,sebastian2020memory}. In particular, there is renewed interest in computing systems based on optics, which could potentially handle matrix-vector multiplication in a more energy-efficient way\cite{rios2019memory,feldmann2021parallel,wang2022optical,spall2020fully,miscuglio2020massively,zhou2021large,mcmahon2023physics}. Despite promising initial results, developing a highly parallel, programmable, and scalable optical computing system capable of rivaling electronic computing hardware still remains elusive. In this context, we propose a hyperspectral in-memory computing architecture that integrates space multiplexing with frequency multiplexing of optical frequency combs and uses spatial light modulators as a programmable optical memory, thereby boosting the computational throughput and the energy efficiency. We have experimentally demonstrated multiply-accumulate operations with higher than 4-bit precision in both matrix-vector and matrix-matrix multiplications, which suggests the system's potential for a wide variety of deep learning and optimization tasks. This system exhibits extraordinary modularity, scalability, and programmability, effectively transcending the traditional limitations of optics-based computing architectures. Our approach demonstrates the potential to scale beyond peta operations per second, marking a significant step towards achieving high-throughput energy-efficient optical computing.
}}

The recent breakthroughs in machine learning, exemplified by AI models such as ChatGPT, have revolutionized myriad industries such as healthcare, finance, retail, automotive, and manufacturing\cite{krizhevsky2012imagenet,esteva2019guide,jumper2021highly,heaton2017deep,wang2018deep,vaswani2017attention}. These transformations have led a surge in demand for extensive matrix-vector multiplication (MVM) operations that are essential for a wide range of algorithms, especially in deep learning\cite{lecun2015deep} and large-scale optimization. This growing computational requirement has presented challenges for traditional von Neumann computing architectures, driving researchers to explore alternate approaches, such as in-memory computing\cite{tsai2018recent,sebastian2020memory}. By incorporating non-volatile memory elements within the processor, in-memory computing effectively addresses the performance bottleneck arising from the separation of memory and processing units in conventional architectures. This leads to more efficient data movement, reduced power consumption, and the capability for highly parallel computations. Numerous in-memory computing systems have been demonstrated using analog circuits that utilize resistive switching\cite{ielmini2018memory,ambrogio2018equivalent,sheridan2017sparse,jung2022crossbar,le2018mixed}. While such analog devices (e.g. memristor crossbar arrays) can efficiently sum partial products using Kirchhoff’s current law\cite{yao2020fully}, they often face limitations in terms of speed, energy efficiency, and precision due to challenges such as parasitic capacitance, inherent resistance, and current leakage\cite{xiao2020analog}.

\begin{figure}[b]
  \centering
  \includegraphics[width=8.5cm]{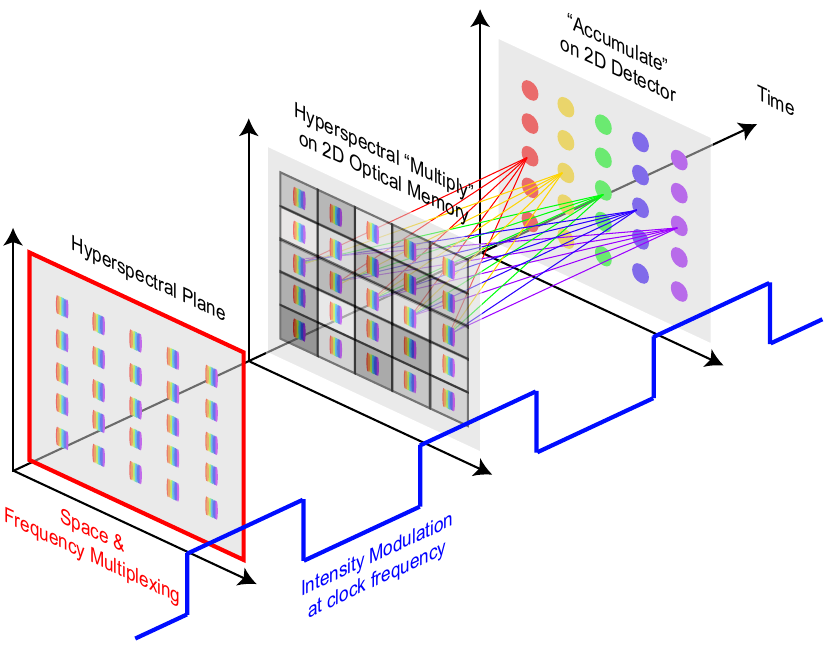}
\caption{
\textbf{Concept of Hyperspectral In-Memory Computing.} 
Boosting computational throughput by combining space and frequency multiplexing with intensity modulation at the computational clock frequency. In this approach, the weight matrix is stored in a 2D optical memory, allowing for efficient multiply-accumulate (MAC) operations as the hyperspectral plane progresses through the system.}
\label{fig:fig1}
\end{figure}

\begin{figure*}[hbtp]
  \centering
  \includegraphics[width=17cm]{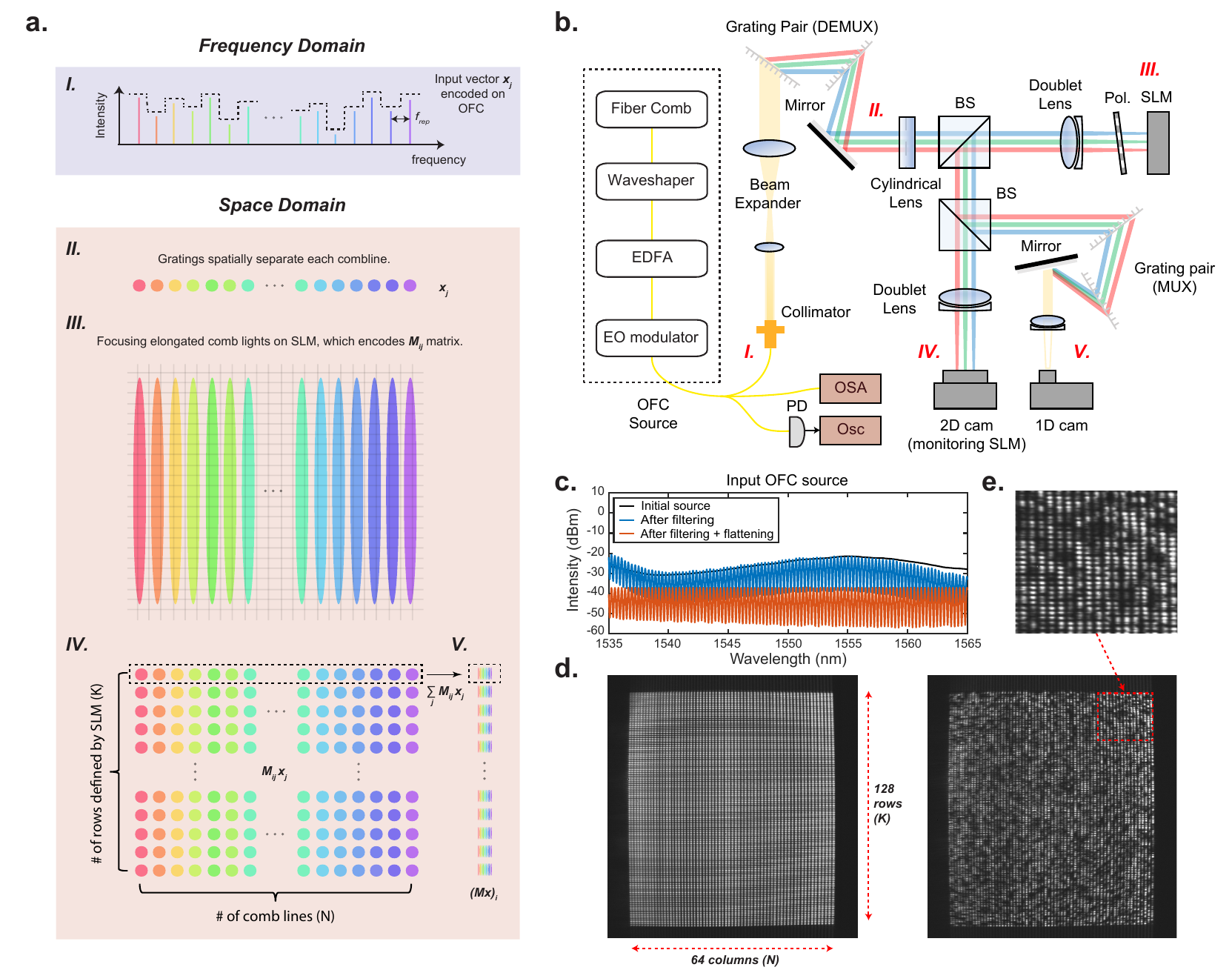}
\caption{
\textbf{Optical Matrix-Vector Multiplication via Mixed Space-Frequency Multiplexing: Operating Principles and Experimental Setup.}
(a) Illustration of optical matrix-vector multiplication (MVM) via mixed space-frequency multiplexing of optical frequency combs (OFCs). I: input vector $x_{j}$ is encoded on OFC by Waveshaper. II: Grating spatially separate each combline. III \& IV: Focusing elongated comb lights on spatial light modulator (SLM), which defines matrix elements $M_{ij}$ and performs element-wise multiplication by attenuation. V: Final result $(M \times x)_{i}$ is detected by 1D IR camera. (b) Experimental setup: The input OFC source, characterized by a large frequency spacing set by the Waveshaper and a lower pulse repetition frequency from the EO intensity modulator, is introduced into the free-space setup. Each pulse carries the information of a vector and performs single-shot matrix-vector multiplication as it passes through the SLM, and is then detected at the line-scan camera. The modulated input OFC source is also detected and characterized using an optical spectrum analyzer (OSA), fast photodetection (PD), and an oscilloscope (Osc). EO modulator: Electro-Optic modulator; EDFA: Erbium-Doped Fiber Amplifier; BS: Beam Splitter; Pol.: Polarizer; MUX: Multiplexer; DEMUX: Demultiplexer. (c) Typical optical spectra of the input OFC source, captured before and after spectral filtering and flattening. (d) 2D images of matrices loaded onto SLM: Uniform (left panel) and random 4-bit matrix (right panel) (e) Zoomed-in image of the random matrix showing the matrix elements encoded with 4 bits.
}
\label{fig:fig2}
\end{figure*}

Concurrently, there is a resurging interest in harnessing optics in computing systems to enable energy-efficient MVM, as optics are inherently suited for mathematical operations that are parallel in nature\cite{caulfield2010future,mcmahon2023physics}. Various optical MVM systems have been proposed and demonstrated\cite{rios2019memory,feldmann2021parallel,wang2022optical,spall2020fully,miscuglio2020massively,zhou2021large} since they first emerged several decades ago\cite{goodman1978fully,tamura1979two,athale1982optical,psaltis1985optical}. These systems include those using fiber optics\cite{xu202111}, planar integrated optical devices\cite{rios2019memory,prabhu2020accelerating, zhou2023memory, meng2023compact}, and three-dimensional optical systems that utilize freespace optics\cite{wang2022optical,spall2020fully,miscuglio2020massively,zhou2021large, lin2018all, zuo2019all, zhou2020situ}. Systems based on fiber optics and optical communication technology have demonstrated high computational throughput thanks to fast modulation and wavelength division multiplexing\cite{xu202111}. However, the serial information encoding consumes computational time, and scaling the system appears challenging. Planar integrated optical devices have shown interesting small-scale demonstrations, but large-scale integration of various devices, such as optical waveguides, programmable optical memory elements, and electronic control circuits, within a limited two-dimensional space is not straightforward\cite{feldmann2021parallel}. Three-dimensional optical systems based on free-space optics, which benefit from scalable display, imaging, and camera technology, hold great potential. Yet, the systems demonstrated so far only utilize space multiplexing, and the frequency dimension remains largely unused\cite{wang2022optical,bernstein2023single,spall2020fully}. Consequently, compared to advanced electronic digital computers, the development of an optical computing system that exhibits high parallelism, precise programmability, and scalability continues to be an ongoing challenge. Here, we propose and demonstrate an optical computing system that incorporates all of these features, drawing inspiration from parallel information processing schemes found in diverse technological domains, including optical communications\cite{winzer2018fiber}, spectroscopy\cite{diddams2007molecular}, imaging\cite{bao2019imaging}, and display technologies\cite{joo2020metasurface}. In our system, computations are primarily executed on a two-dimensional spatial light modulator (SLM)\cite{efron1994spatial,weiner2000femtosecond,SantecSLM}, functioning as a programmable optical memory, where the weight matrix either remains static or experiences slow updates. This setup allows optics to predominantly handle energy-efficient and parallel data movement, while electronics ensure the system's programmability. Importantly, our system architecture leverages scalable optical computing methods and components to substantially increase computational throughput. This strategy includes the integration of hyperspectral operations\cite{plaza2009recent} using optical frequency combs (OFCs)\cite{diddams2020optical,fortier201920} that, with coherently-locked massively parallel laser comb lines, could facilitate both frequency and space multiplexing (See Figure \ref{fig:fig1}).

\begin{figure*}[hbtp]
  \centering
  \includegraphics[width=17cm]{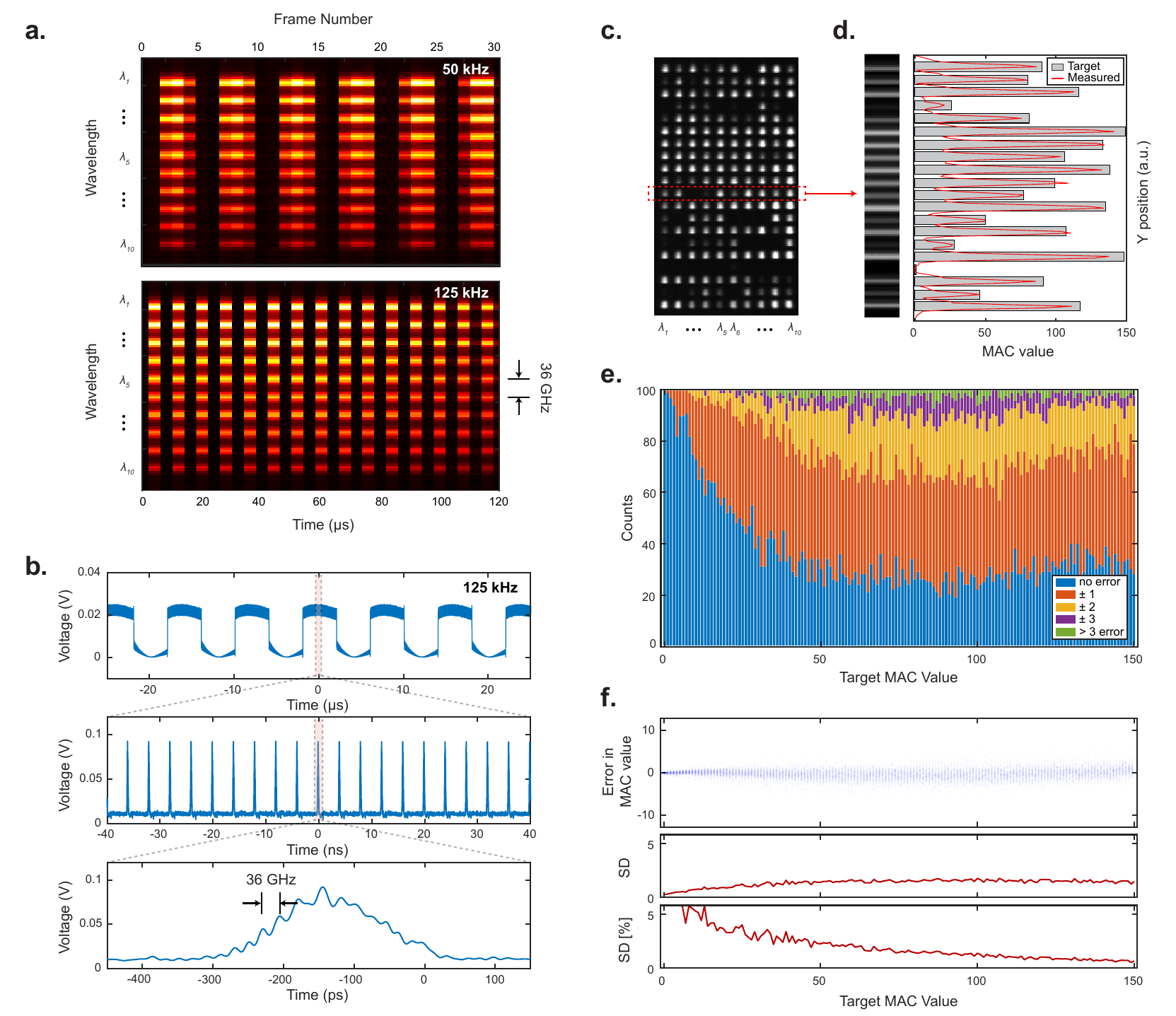}
\caption{
\textbf{Single-Shot Matrix-Vector Multiplication through Intensity Modulation: Experimental Details and Error Analysis} (a) Time evolution of line-scan camera image at the frame rate of 250 kHz, which shows the intensity modulation of input OFC source with 10 frequency components separated by 36 GHz. The intensity modulation rate is 50 kHz (top) and 125 kHz (bottom). (b) Fast photodetection of the input OFC source shows the intensity modulation rate of 125 kHz (top), the fiber comb repetition rate of 250 MHz (middle), and the Waveshaper filtering FSR of 36 GHz (bottom). (c) 2D camera image of an optical matrix after element-wise multiply operation. (d) Line-scan camera image after multiply-accumulate operation (left panel). The measured MAC values show good agreement with the target MAC values (right panel). (e) Error distribution across all possible MAC values, utilizing a unit input vector and a 4-bit random matrix. For each MAC value, 100 operations are conducted for analysis. (f) Absolute error at each target MAC value (Top panel) and the standard deviation (SD) of the error distribution (Middle panel) and as a percentage (Bottom panel).
}
\label{fig:fig3}
\end{figure*}

\begin{figure*}[hbtp]
  \centering
  \includegraphics[width=17cm]{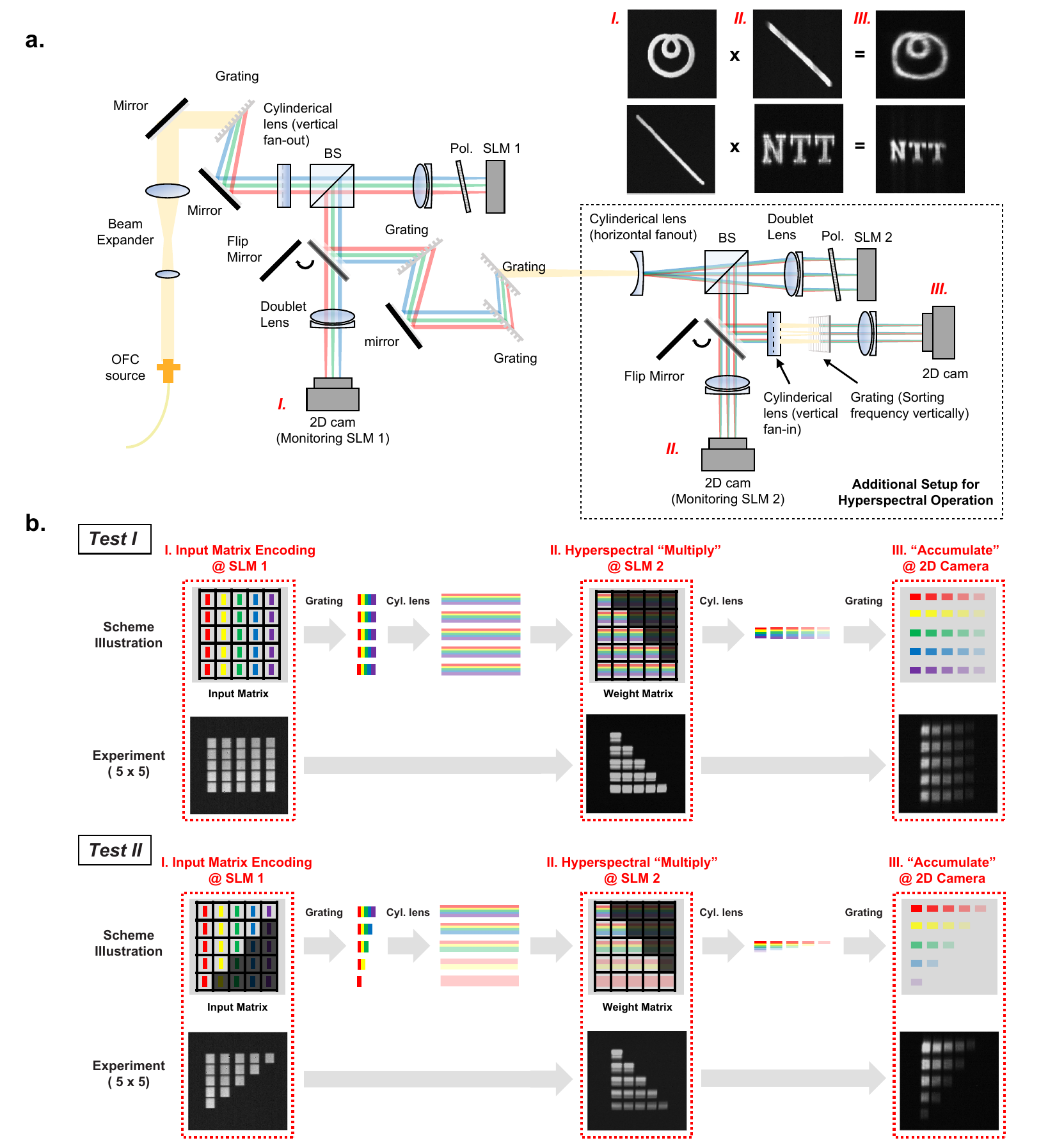}
\caption{
\textbf{Hyperspectral Multiply-Accumulate} (a) Extended experimental setup for the open-loop hyperspectral multiply-accumulate (MAC) operation. An additional configuration (outlined in the dotted box) complements the matrix-vector multiplication (MVM) setup, enabling matrix-matrix multiplication (MMM) through the hyperspectral multiply-accumulate process. Matrices are encoded onto SLM 1 and SLM 2, and the resultant matrix is captured by a 2D camera. As examples, encoding the NTT logo and identity matrix (I and II), each with an approximate size of 100 by 100, yields an output matrix displaying the NTT logo (III). (b) Illustrations of MMM performed through hyperspectral operation are paired with images from two test experiments. Both the theoretical illustrations and experimental results demonstrate strong agreement, highlighting the ``Hyperspectral Multiply-Accumulate'' process effectively captured in the 2D camera images. Test I shows the multiplication of an all-ones matrix with a lower triangular matrix, while Test II demonstrates the multiplication of two triangular matrices. For clarity, we've presented straightforward examples with hyperspectral factors of 5. Operational principles, as well as experimental results for higher hyperspectral factors, are detailed in the Supplementary Section IV. Note: Distortions in the images are mainly due to accumulated optical alignment errors.
}
\label{fig:fig4}
\end{figure*}

In the experiment, we first demonstrated an optical MVM system utilizing mixed space-frequency multiplexing of OFCs\cite{hase2018scan,bao2019imaging}, which is a groundwork for our hyperspectral in-memory computing system. By connecting frequency and space dimension through dispersive elements, the multiplexing method enables all-to-all interaction among multiple frequency channels via a programmable SLM while maintaining the parallel data transfer. For the input source, a fiber OFC in the optical C-band with a 250 MHz pulse repetition rate is coarsely filtered to achieve 36 GHz spacing (see Methods). This effectively generates OFCs with a 36 GHz frequency spacing and an intensity modulation of 250 MHz (see Figure \ref{fig:fig2}c for typical optical spectra). Then, the input vector element ($x_j$) is encoded in the intensity of each 36 GHz comb line using line-by-line waveshaping\cite{weiner2000femtosecond}(see Figure \ref{fig:fig2}a I). Subsequently, the input optical source with an average power of approximately 1 mW is introduced to the system illustrated in Figure \ref{fig:fig2}b. The comb lines are spatially separated by two gratings (see Figure \ref{fig:fig2}a II), fanned out vertically using a cylindrical concave lens, and then focused onto the SLM (see Figure \ref{fig:fig2}a III). The SLM encodes the matrix elements ($M_{ij}$) as attenuation weights for multiplication with the input vector. We employed doublet lenses and a beam expander to minimize aberration and image plane distortion on the SLM surface. A linear polarizer was utilized to operate the phase-only SLM for intensity modulation. Here, SLM is used as a programmable optical memory due to its precision and scalability. While our system uses intensity modulation for encoding information, it is worth mentioning that other optical data communication techniques, like amplitude modulation, phase shift, or frequency shift, can also be applied similarly, owing to the coherence of the input OFC source.

Using a beam splitter (BS), a two-dimensional short-wavelength infrared (SWIR) camera captures the resultant optical matrix ($M_{ij}x_j$) from the SLM (see Figure \ref{fig:fig2}a IV). Figure \ref{fig:fig2}d shows the images of two example matrices on the SLM, revealing distinct elements within a 64-pixel by 128-pixel matrix (see Figure \ref{fig:fig2}e). Finally, the same optical matrix is summed up in the horizontal direction and its optical intensity ($\sum_{j} M_{ij}x_j$) is detected by a line-scan SWIR camera, completing the matrix-vector multiplication, also known as the multiply-accumulate (MAC) operation (see Figure \ref{fig:fig2}a V). Here, the total insertion loss of the freespace part of the setup is approximately 12 dB. The system non-uniformity, including the Gaussian intensity profile on the SLM surface, is detected using the line-scan camera and calibrated by adjusting the phase of the SLM pixels (see Methods and Supplementary Section II).

To verify the system operation, the input OFC source, after encoding an arbitrary 10 $\times$ 1 vector, is intensity-modulated and a unit diagonal matrix is loaded on the SLM, which allows us to monitor the frequency-resolved time evolution of the input pulse stream. Figure \ref{fig:fig3}a shows the time evolution of input pulse stream at 50 kHz and 125 kHz modulation, and each optical pulse contains ten frequency components separated by 36 GHz. Here, the detectable modulation rate is limited by the frame rate of the line-scan camera (250 kHz). In principle, pulse mode operation at the fiber comb repetition rate (250 MHz) or even higher modulation rate is also feasible if the readout from each photodetector pixel is parallelized. Fast photodetection of the input OFC source clearly shows the three modulation frequencies (125 kHz, 250 MHz, and 36 GHz) of the input OFC source (see Figure \ref{fig:fig3}b). With 250 MHz modulation and the demonstrated matrix size of 128 by 64, the current system already demonstrates a computational throughput exceeding 2 tera operations per second (TOPS). 

\begin{figure*}[hbtp]
  \centering
  \includegraphics[width=17cm]{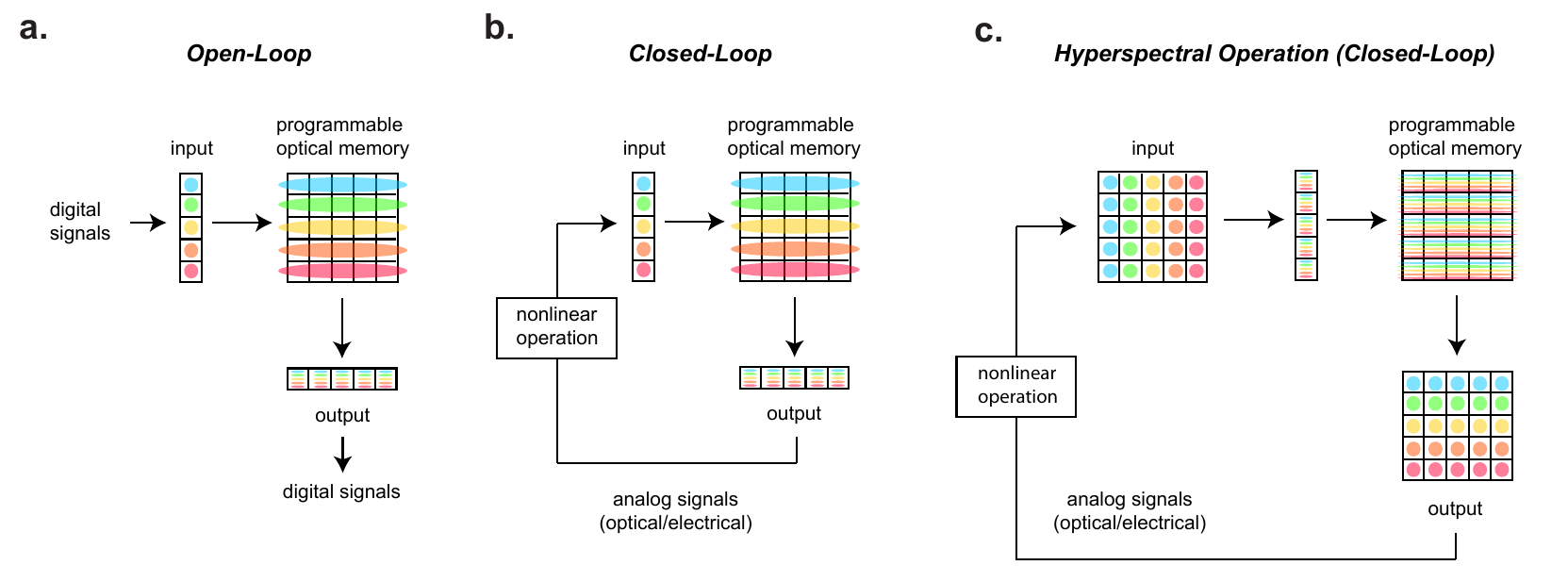}
\caption{
\textbf{Architectures of optical in-memory computing systems} (a) Open-loop optical MVM system. (b) Closed-loop optical MVM system operating as a physical solver for optimization problems. (c) Closed-loop optical MMM system that utilizes both space and frequency multiplexing via hyperspectral operation. Note: Different colors indicate different optical frequencies. Single-frequency lasers or incoherent light sources can replace the optical frequency comb in systems (a) and (b) if only space multiplexing is required. Although the analog signal pathways in diagrams (b) and (c) support parallel data transmission, a single line is shown for simplicity.}
\label{fig:fig5}
\end{figure*}

To assess the computational accuracy of our system, we analyzed the error distribution for every potential MAC value. When a unit input vector was mapped onto the OFC and a 20 by 10 random matrix was loaded onto the SLM (see Figure 3c), the optical summation outcomes (MAC values) were captured using the line-scan camera (see Figure \ref{fig:fig3}d). Here, the matrix was encoded using only non-negative weights with 4-bit, but encoding negative weights is feasible\cite{spall2020fully}. While higher bit encoding is achievable, 4-bit precision is often sufficient for synaptic operations in most optimization or deep learning tasks to attain ``software-equivalent'' accuracy\cite{okazaki2022analog} (see Supplementary Section III). We conducted 100 measurements for each target MAC value, ranging from 0 to 150 (see Figure \ref{fig:fig3}e). As the target MAC values increase, the standard deviation of the error tends to grow, reaching a saturation point after a certain value (as shown in Figure \ref{fig:fig3}f, top and middle panels). Accordingly, the relative error, defined as $\lvert$measured MAC value - target MAC value$\rvert$/(target MAC value), sees its standard deviation decrease to below 2 percent as the target MAC value rises. These errors can be attributed to factors like intensity fluctuations of the input OFC source, crosstalk between adjacent pixels, and optical alignment errors. We expect the standard deviation of the relative error to remain consistent at a similar level, even when the system operates with a larger matrix size or at higher bit encoding. It's worth noting that noise (i.e. errors in the MAC value) up to a certain threshold might not significantly impact the computational results\cite{rasch2023hardware}. We verified this by analyzing the classification of MNIST data, a basic yet illustrative example, under different noise levels (see Supplementary Section III). Furthermore, in certain optimization tasks, noise can even help the system escape from a local optimum solution\cite{wang1999training,goh2007investigation}.

In the first experiment, the system operates in an open-loop, whereby the encoding of the input vector and the readout of output MAC results are executed independently using commercial digital electronic circuits (see Figure \ref{fig:fig5}a). For high-throughput computation in open-loop operations, fast external modulation and readout are essential. Electro-optical modulators or digital micromirror devices can be utilized for fast input vector encoding, while a parallel readout electrical circuit is needed for each photodetector pixel. When the system operates in closed-loop mode with a desired nonlinear operation (see Figure \ref{fig:fig5}b), it can function as a physical solver for optimization problems (e.g., the Ising problem, quadratic unconstrained binary optimization) without requiring fast external modulation and readout. In such iterative systems, computation can be performed entirely in an analog manner, with only the initial problem loading and the final solution readout managed using digital electronics. This hybrid architecture combines the speed of analog computing with the flexibility of a digital interface\cite{mourgias2023analog}.
It's worth noting that for enabling fast pixel-by-pixel parallel modulation in the closed-loop system, it may be essential to introduce a novel device that enables a direct one-to-one connection between the photodetector pixel and the modulator pixel, rather than relying on a camera and SLM connected through a serial bus. This ensures that both data transfer and computation remain fully parallel throughout the computational cycle, avoiding the typical computational slowdowns arising from bottlenecks due to the conversion of information from parallel to serial format\cite{igehy1998design,schikore2000high,montrym1997infinitereality}.
\begin{table*}[hbtp]
\centering
\caption{\bf Estimated System Performance }
\begin{tabular}{cccccc}
\hline
 & Current \hspace{2mm} &  Near-term \hspace{2mm} & Long-term \\
 & (open-loop) \hspace{2mm} & (closed-loop) \hspace{2mm} & (closed-loop)\\
\hline
Matrix Size \hspace{2mm} & 128 $\times$  64 \hspace{2mm} & 300 $\times$  300 \hspace{2mm} & 1000 $\times$  1000 \\
Hyperspectral Factor \hspace{2mm}   & $\times$ 1 ($\times$  10) \hspace{2mm} & $\times$  30 \hspace{2mm} & $\times$  100 \\
Clock Frequency  \hspace{2mm}   & 250 MHz$^{\dagger}$ \hspace{2mm} & 1 GHz \hspace{2mm} & 1 GHz \\
Computational Throughput \hspace{2mm}  & 2.048 TOPS (20.48 TOPS) \hspace{2mm} & 2.7 PetaOPS \hspace{2mm} & 100 PetaOPS \\
Total Power Consumption$^{\dagger\dagger}$  \hspace{2mm}  & 11.9 W \hspace{2mm} & 27.7 W \hspace{2mm} & 206 W \\
Power Efficiency \hspace{2mm}  & 5.8 W/TOPS \hspace{2mm} & 10.26 W/PetaOPS \hspace{2mm} & 2.06 W/PetaOPS \\
\hline
\end{tabular}\\
\raggedright
\vspace{3mm}
\hspace{15mm} $^{\dagger}$ We assume an external modulation and readout speed of 250 MHz.\\
\hspace{15mm} $^{\dagger\dagger}$ Details of the power consumption estimation are discussed in the Supplementary Information.\\
  \label{tab:table1}
\end{table*}

Our optical MVM system in the first experiment can achieve a computational throughput of 2.048 TOPS and has the potential to approach Peta operations per second (PetaOPS) by increasing the clock frequency and expanding the matrix size. However, to surpass PetaOPS, hyperspectral operation is essential, as it leverages the full parallelism of OFCs (see Figure \ref{fig:fig5}c and Table \ref{tab:table1}). In the follow-up experiment, we extended our initial optical MVM system by incorporating the hyperspectral operation setup, as indicated by the dotted box in Figure \ref{fig:fig4}a. We performed matrix-matrix multiplication (MMM) operations by utilizing hyperspectral encoding, where each SLM pixel encodes a matrix weight across multiple comb lines. This is fundamentally a batch processing of matrix-vector multiplication using wavelength-division multiplexing \cite{tamura1979two}. We carried out numerous MMM tests, and the results were consistent with theoretical predictions. This includes the multiplication of the NTT logo with the identity matrix, as illustrated in Figure \ref{fig:fig4}a. To clarify the hyperspectral operation, we presented two example tests with a hyperspectral factor of 5, where each SLM pixel encodes a matrix weight over 5 comb lines (see Figure \ref{fig:fig4}b). Yet, with slight adjustments to our current system, achieving a hyperspectral factor of 10 or even higher was possible (see Figure S6). We also performed the error analysis of MAC values, and the noise remains under 5 percent regardless of matrix size (see Figure S7). This observed noise level is slightly higher compared to our initial MVM experiment, likely due to increased insertion loss of the free-space setup and compounded optical alignment errors. It's worth noting that Figure \ref{fig:fig4} (along with Figure S5) illustrates one particular method for implementing hyperspectral operations. Other types of hyperspectral encoding or hyperspectral decoding at detector pixels\cite{chang2003hyperspectral} are also feasible.

Looking ahead to the near-term system which will work in a closed-loop setting, improvements like better alignment and wider spectral bandwidth allowing for larger matrix sizes (300 $\times$ 300), greater hyperspectral factor ($\times$ 30) and increased modulation speed (1 GHz) should facilitate a performance of up to 2.7 PetaOPS with an estimated power efficiency of 10.26 W/PetaOPS. In the long run, targeting a performance of 100 PetaOPS with a hyperspectral enhancement of $\times$ 100 and a matrix size of 1000 $\times$ 1000, we project that the future system might achieve 100 PetaOPS with a power efficiency close to 2 W/PetaOPS. This is a two orders of magnitude efficiency boost compared to state-of-the-art electronic GPUs (see Table I and Supplementary Section V for additional details on system performance estimation). Furthermore, development of multispectral SLMs\cite{mansha2022high} could further enhance the throughput and adaptability of such a hyperspectral in-memory computing system.


Our proposed optical computing system leverages the dimensions of frequency, space, and time to maximize computational throughput and power efficiency. The design prioritizes scalability, merging space and frequency multiplexing with scalable SLM and OFC technologies. Given the significant efforts from both industry and academia to improve SLM capabilities\cite{lazarev2019beyond}, driven by their expansive applications in optical imaging, free-space optical communications, and augmented/virtual reality (AR/VR), our modular system can benefit directly from these advancements in SLM speed\cite{panuski2022full} and resolution\cite{mansha2022high}. Simultaneously, the progress in OFC technology\cite{diddams2020optical,fortier201920} further complements our system's capabilities. This strategy of modular system design not only allows for individual component advancements but also fosters overall system performance enhancement. Additionally, the system can be scaled further by incorporating two-dimensional arrays of optical elements such as lens or SLM arrays. We can also tap into polarization multiplexing\cite{deng2020malus} for an added boost in computational throughput. The integration of next-generation innovations like metalenses\cite{chen2018broadband}, chip-integrated OFCs\cite{xiang2021laser}, and amplifiers\cite{liu2022photonic} hints at the potential for significant miniaturization. Such progress could set the stage for our system to be seamlessly integrated into data centers as a rack-mounted solution. As these component technologies continue to evolve, our proposed architecture could lead to a new era of energy-efficient optical information processing, possibly exceeding PetaOPS in future cloud computing environments.


\begin{footnotesize}

\section*{Methods}
\noindent \textbf{Experimental apparatus details}
During our preliminary experiment, we used an electro-optically modulated optical frequency comb (OFC) with a 45 GHz spacing as an optical source. This frequency spacing was sufficiently large for spatial separation via dispersive elements. For enhanced stability, a fiber OFC source with a pulse repetition rate of 250 MHz (Menlo Systems FC1500-250-ULN) was used as an alternative input source, after spectral filtering on a 36 GHz FSR grid using a Waveshaper (see Figure \ref{fig:fig2}c). To improve energy efficiency, it is also possible to use a microcomb with a large FSR for frequency-multiplexing and to apply lower frequency intensity modulation to it. We used optical sources and other components for optical C-band due to the availability of equipment in our laboratory. For the purpose of a programmable optical memory, we utilized a spatial light modulator (Santec SLM-200) with a resolution of 1920 $\times$ 1200 pixels and a frame rate of 120 Hz. In our experiments, we grouped multiple adjacent pixels together to form an 'effective pixel'. This grouping typically ranged from 1 $\times$ 1 pixel to 30 $\times$ 30 pixels. The convex spherical lenses were achromatic doublets with a focal length of 75.4 mm. The gratings were transmission type with 1000 lines/mm. To capture the weight matrix encoded on the SLM, we used an InGaAs 2D camera (Pembroke Instruments SenS 1280), which has a resolution of 1280 $\times$ 1024 pixels and a maximum frame rate of 60 Hz. To detect the MAC value, we opted for the InGaAs line-scan camera (Xenics Manx 2048 R CXP 260), which offers 2048 pixels and a frame rate up to 256 kHz. Faster cameras with a larger number of pixels are readily available thanks to advanced CMOS technology if the demonstrated system is operated at wavelength ranges below 1 micron. To characterize the input pulse using fast photodetection, we used a 100 GHz photodetector (Finisar XPDV4121R-WF-FA) and analyzed the data with a 256 GSa/s realtime oscilloscope (Keysight UXR-Series).\\

\noindent \textbf{System calibration}
Several calibration processes are required before the operation of both MVM and MMM systems. The non-uniformity of the optical system, including the Gaussian intensity profile on the surface of the 2D spatial light modulator (SLM), is detected using a line-scan IR camera. This system non-uniformity is corrected by adjusting the phase of the SLM pixels (see Figure S2). In addition, the non-uniform and non-linear mapping (see Figure S1) between the phase and intensity of each SLM pixel is calibrated when the problem weight matrix is loaded onto the SLM. The intensity of the input OFC is adjusted so that the IR camera pixels do not saturate and operate within the linear response range.\\

\noindent \textbf{Post-processing of images for MAC results} 
After loading the target input vector and matrices into the system, the MAC results are obtained from the images captured by the camera. To remove the camera's DC offset and any unintended background illumination, a background image (captured with the SLM transmission set to zero) is subtracted from the image frames containing the MAC results. Subsequently, the final MAC results are determined by comparing the intensities of each captured frame with a reference `normalization' frame, which is created by loading all-1 matrices onto the SLMs.

\medskip

\end{footnotesize}

\bibliography{main}

\section*{Acknowledgements}
We thank Tianyu Wang, Edwin Ng, and Satoshi Kako for helpful discussions. 

\section*{Author Contributions}
M.G.S originated the concept, and all authors collaborated in formulating the ideas for the experimental demonstration. M.H. and B.J.S. conducted the experiment with the assistance from M.G.S. All authors analyzed the data. All authors contributed to writing the manuscript. Y.Y and M.G.S supervised the project.

\vbox{}
\noindent \textbf{\large Competing Interests} \\ 

\vbox{}
\noindent \textbf{\large Author Information} \\

\vbox{}
\noindent \textbf{\large Data Availability} \\ The data that support the plots within this paper and other findings of this study are available from the corresponding authors upon reasonable request.

\vbox{}
\noindent \textbf{\large Code Availability} \\ The code that support the plots within this paper and other findings of this study are available from the corresponding authors upon reasonable request.

\ifarXiv
    

\section*{Supplementary material}

# Supplementary Information for
# "Hyperspectral In-Memory Computing with Optical Frequency Combs and Programmable Optical Memories"

Mostafa Honari Latifpour[1,2,†], Byoung Jun Park[1,3,†], Yoshihisa Yamamoto[1], Myoung-Gyun Suh[1,*]

[1]Physics & Informatics Laboratories, NTT Research, Inc., Sunnyvale, CA 94085, USA
[2]The Graduate Center, City University of New York, New York, NY 10016, USA
[3]KU-KIST Graduate School of Converging Science and Technology, Korea University, Seoul 02841, Republic of Korea
[†] These authors contributed equally to this work.
[*]Corresponding author: Myoung-Gyun Suh (email: myoung-gyun.suh@ntt-research.com)

**This PDF file includes:**

FIG. S1 - S8
TABLES S1 - S2

## I. SPATIAL LIGHT MODULATOR AS OPTICAL MEMORY

The spatial light modulator (SLM) used in this work is Liquid Crystal on Silicon (LCoS) type. LCoS SLMs utilize birefringent liquid crystals that have the ability to modulate the refractive index of individual pixels when a voltage is applied. This change in refractive index induces a phase shift in the light passing through the liquid crystal. By appropriately placing a polarizer in front of the SLM, this phase modulation can be translated into amplitude modulation. LCoS devices generally have a reflective operation, wherein light is modulated upon reflection off the silicon backing, which distinguishes them from transmissive liquid crystal SLMs.

In our experiments, we use SLMs as optical memories, encoding matrix weights as amplitude attenuation coefficients. The SLM has a resolution of 1280 × 1920 pixels and provides 10-bit precision (that is 1024 levels) for phase modulation. For a given attenuation target, the phase level that yields the desired attenuation is set on the SLM. Hence, an accurate conversion from attenuation to phase is critical for the accurate multiplication operation. FIG. S1 shows the relationship between reflection and SLM phase over the full phase range of the SLM. The highlighted green section serves as a look-up table to determine the phase corresponding to the desired attenuation.

In the context of closed-loop hyperspectral systems described in the main text, it becomes important to achieve significantly higher modulation speeds for each pixel of the SLM (e.g. 1 GHz). LCOS SLMs, constrained by the slow response time of their liquid crystal cells, typically operate at modulation speeds of less than 1 kHz, making them unsuitable for fast modulation within closed-loop systems. Consequently, alternative modulation mechanisms, such as those based on changes in free-carrier density[1] or material phase[2], appear to be a requisite. An array of vertical cavity surface-emitting lasers (VCSELs) can also be used for optical modulation, as well as serving as the input optical source or optical amplifier[3].

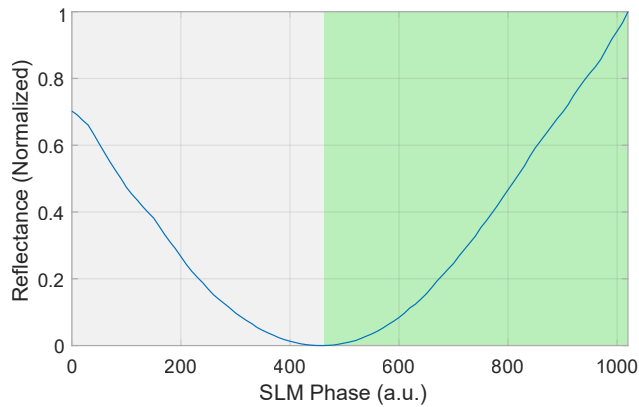

FIG. S1. The reflection from the SLM, as a function of the pixel phase, is measured at a wavelength of 1550 nm. The area highlighted in green indicates the target attenuation level being encoded.

## II. CALIBRATION OF SYSTEM NON-UNIFORMITY

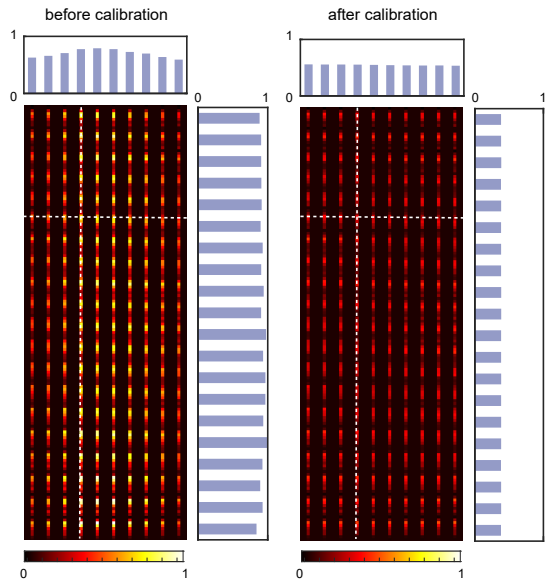

FIG. S2. **Calibration of system non-uniformity** Horizontally-stacked line-scan camera images showing 10 × 20 unit matrix before (left) and after calibration (right).

## III. ANALYSIS OF ERRORS IN MAC VALUES

In the main text, we presented experimental data with up to 4-bit encoding, where each matrix element can encode 16 discrete weight levels. We focused on 4-bit as it is sufficient for many applications. Nonetheless, our system is capable of operating at higher bit encoding while maintaining a similar relative noise level (see the 5-bit result in FIG. S3). In our current experimental system, the standard deviation of the relative error is below 5 percent, ensuring that the computational accuracy of the MNIST classification task remains robust, as shown in FIG. S4. Furthermore, by reducing the optical alignment errors and minimizing crosstalk between adjacent pixels, we expect the system could achieve reduced overall noise and operate at higher bit-precision. By extending the acquisition duration or averaging results from consecutive measurements taken over time, we might also improve the bit-precision due to a reduction in $\sigma$. However, such adjustments would lead to decreased operation speeds.

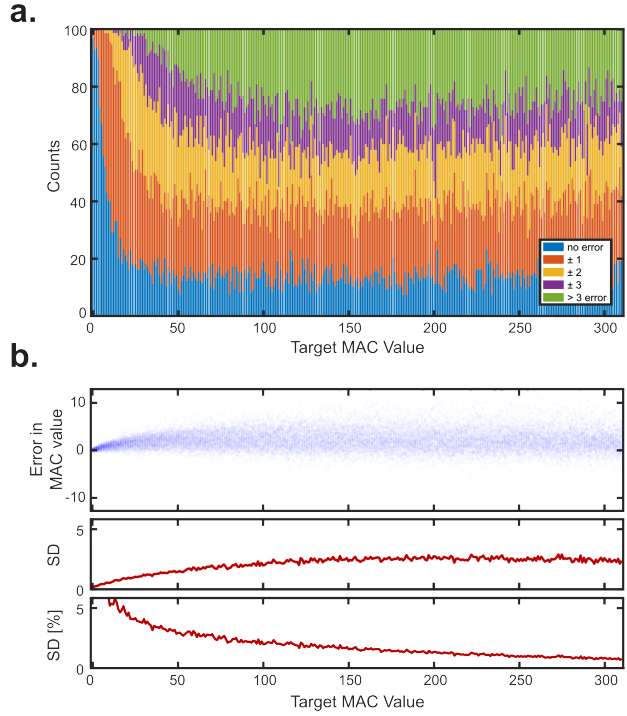

FIG. S3. **Error Analysis of MAC values (5-bit)** (a) Error distribution for each possible MAC value. Unit input vector (1 × 10) and 5-bit random matrix (10 × 10) are used. At each MAC value, 100 MAC operations are performed for the analysis. (b) Absolute error at each target MAC value (Top panel) and the standard deviation (SD) of the error distribution (Middle panel) and as a percentage (Bottom panel).

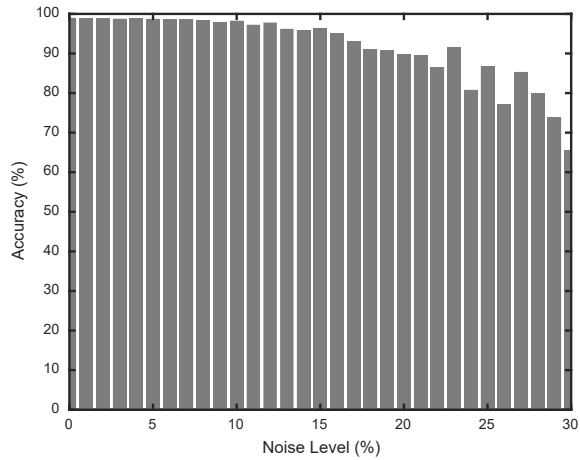

FIG. S4. **Classification of MNIST data with different noise levels.** The classification accuracy of MNIST data is assessed when the LeNet-5 model is trained under different noise levels, denoted by their standard deviations. The training set comprises 60,000 images, while the test set has 10,000 images. Each image has a size of 28 × 28 pixels. The training process spans 10 epochs.

## IV. ADDITIONAL DETAILS OF HYPERSPECTRAL OPERATION

Hyperspectral operations utilize the wideband response of a pixel to process additional information without increasing its physical dimensions. Figure S5 illustrates a method for implementing matrix-matrix multiplication using this hyperspectral approach. The associated experimental setup can be seen in Figure 4a

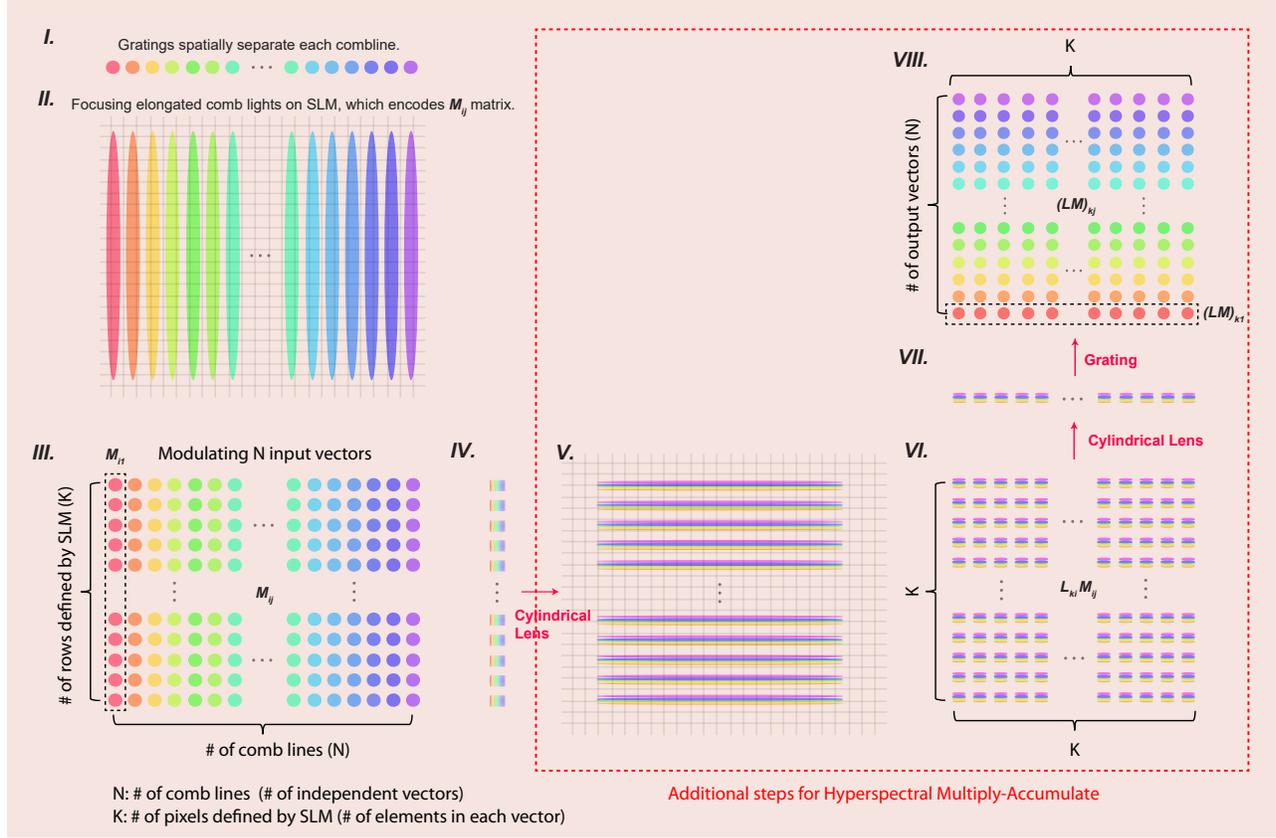

FIG. S5. **Operational Principle of Hyperspectral Multiply-Accumulate.** Matrix-matrix multiplications performed through hyperspectral operation.

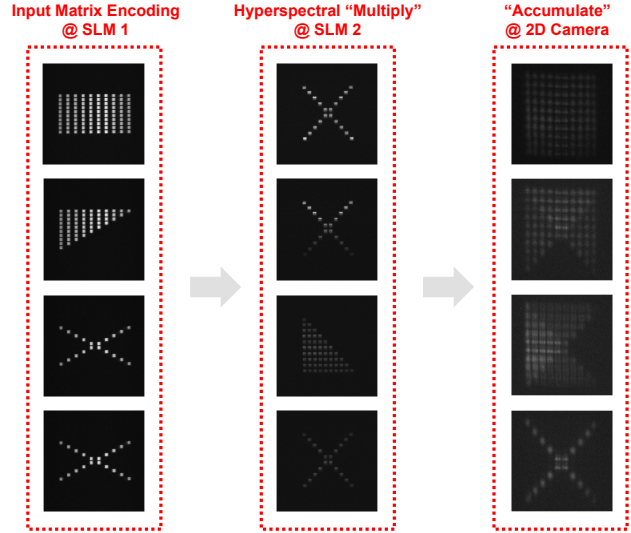

FIG. S6. **Additional Tests of Hyperspectral Multiply-Accumulate.** Matrix-matrix multiplications performed through hyperspectral operation captured in the 2D camera images. Matrix size of 10 × 10 (hyperspectral factor of 10) is used for the additional tests.

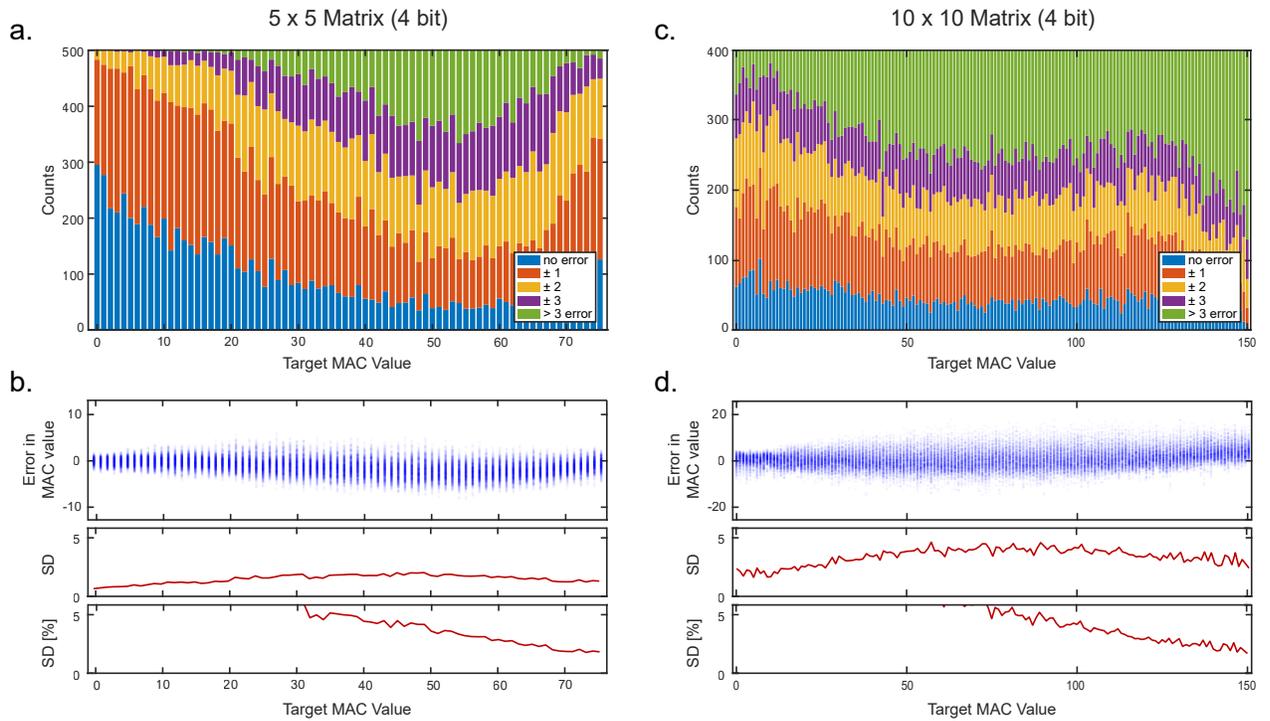

FIG. S7. **Error Analysis of Hyperspectral Multiply-Accumulate** (a) Error distribution for each possible MAC value. All ones input matrix and 4-bit random matrix are used. The size of matrices is 5 × 5. At each MAC value, 500 MAC operations are performed for the analysis. (b) Top: Absolute error per target MAC value. Middle: Standard deviation (SD) of error. Bottom: Error as a percentage. (c) As in (a) but for a 10x10 matrix, analyzing 400 operations per MAC value. (d) Metrics identical to (b). Despite the difference in matrix size, the relative error is consistent. Potential reductions in relative error can be achieved through optimizations, such as improved alignment of free space optics, use of diffraction-limited achromatic lenses, and enhanced calibration.

## V. ESTIMATION OF SYSTEM PERFORMANCE

In this section, we will discuss the preliminary estimation of the system's performance, focusing on the power efficiency as highlighted in Table S2.

1. Current system:

We have estimated the current system's total power consumption based on the actual numbers from our experiments and the specifications of currently available commercial parts known for their low power consumption, not taking into account potential future enhancements to individual components.

A high-level schematic of the MVM system is depicted in FIG. S8. It includes a light source which generates a wideband laser light as the optical carrier, intensity modulator array for encoding the input vector information on separate wavelengths of the laser, a diffractive free-space optics setup for fan-out, multiplication, and fan-in of light, and an InGaAs photodiode array[4] for summation and readout. The total power consumption is the sum of the dissipated optical power plus the electrical power used by the electronic systems. Total optical loss of the system from the input OFC to the photodiodes is estimated to be 15 dB, which includes 3 dB loss of the modulators and 12 dB loss of the free-space optics involving diffractive elements and the SLM. Assuming an input vector size of N, matrix size of $N \times K$, and a bit precision of $N_b$ for the detected signal, the total required power can be obtained from the equation below using the assumptions listed in Table S1 for experimental parameters:

$$P_{N \times K}^{(\text{open-loop MVM})} \approx N \times (P_{DAC} + P_{mod}) + P_{SLM} + K \times \left[ \frac{2^{N_b} I_{th}}{\eta_L \eta_o \eta_{PD}} + P_{TIA} + P_{ADC} \right]$$

This gives $P \approx 11.9 W$ for $N = 64$, $K = 128$ and $N_b = 8$. At the clock frequency of 250MHz, we would have $2.048 \times 10^{12}$ MAC operations per second (see Table S2), giving 5.8 W/TOPS (i.e. 5.8 pJ/MAC). It is worth noting that the power consumption estimation is primarily influenced by the SLM ($P_{SLM}$) control unit, which we approximated based on the product datasheet. However, the actual power consumption should be considerably lower since the SLM serves as a static optical memory in our current setup. Moreover, as the system scales up, the constant power consumption of the SLM becomes negligible.

TABLE S1. **Assumptions for Power Consumption Estimation**

| Symbols | Parameters | values |
|---|---|---|
| $\eta_L$ | Wall-plug efficiency[5–7] | 0.1 |
| $\eta_{PD}$ | InGaAs PD responsivity at 1550 nm[8] | 1.0 $A/W$ |
| $I_{th}$ | Current detection threshold[8] | 15 nA |
| $\eta_o$ | Optical power utilization efficiency | 0.03 |
| $\eta_o^{'}$ | Optical power utilization efficiency for hyperspectral MAC | 0.01 |
| $P_{TIA}$ | Transimpedance amplifier power consumption[9–12] | 1 mW |
| $P_{mod}$ | Electro-optic modulator power consumption[13] | 20 mW |
| $P_{DAC}$ | DAC power consumption[14] | 1 mW |
| $P_{ADC}$ | ADC power consumption[15] | 2 mW |
| $P_{SLM}$ | SLM power consumption[16,17] | < 10 W |

2. Future systems with hyperspectral operation:

To boost the computational throughput further, future optical in-memory computing systems will have to use hyperspectral operations. In this scenario, we are planning for the system to only use specific optical bands (e.g. optical C-band) because of the limitations posed by the optical amplifiers' gain bandwidth and the dispersion seen in free-space optical components. In addition, increasing the number of components for hyperspectral operation will also raise the system's insertion loss. Assuming extra 5 dB loss, the total optical loss from the input OFC to the photodiodes becomes 20 dB and $\eta_o^{'} \approx 0.01$.

For a closed-loop hyperspectral system that relies on analog computation, we can eliminate both the ADC and DAC components from the setup. We can also reduce the bit-precision as the system is immune to certain noise level and it does not require digitization. So, let's assume $N_b = 6$. Moreover, if we adopt a novel device that facilitates a direct one-to-one connection between the photodetector pixel and the modulator pixel, as discussed in the main

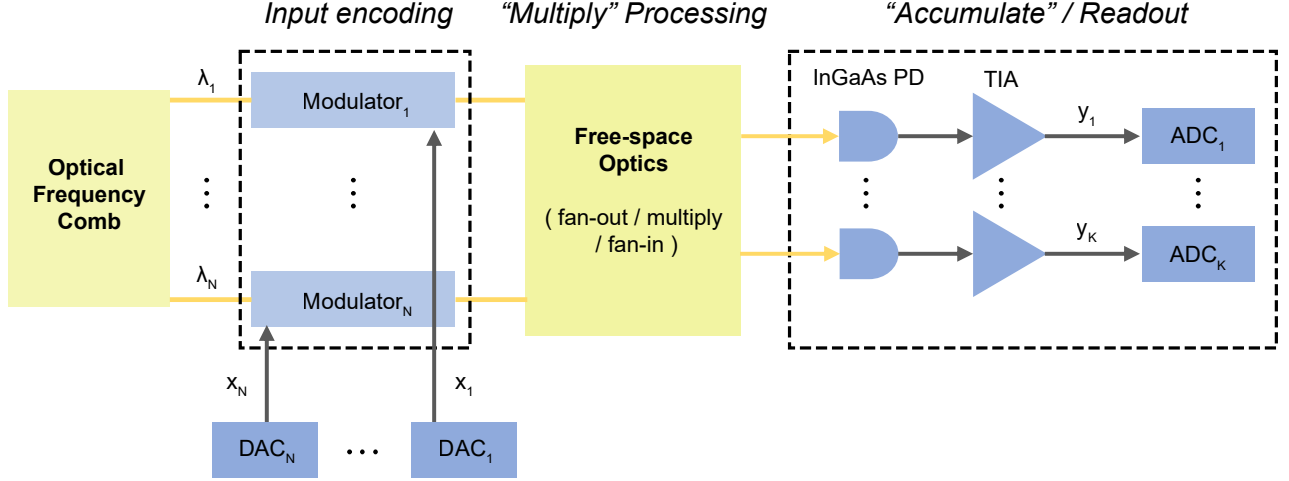

FIG. S8. **High-level schematic diagram of the open-loop optical MVM system used for estimating power consumption.** Power requirements for the closed-loop MMM system with hyperspectral operation are also derived from a closely related, though modified, configuration.

text, the optical power detected at each photodetector pixel will amplify and directly modulate its corresponding modulator pixel. In this case, only one intensity modulator will be need for input optical pulse stream generation. When considering the hyperspectral factor H (here, N = H) for the multiplication of two matrices with size (H × K) and (K × K), the equation for total power consumption is as follows:

$$P_{H \times K \times K}^{(\text{closed-loop MMM})} \approx P_{mod} + P_{SLM} + (H \times K) \times \left[ \frac{2^{N_b} I_{th}}{\eta_L \eta_o' \eta_{PD}} + P_{TIA} \right].$$

This gives $P \approx 27.7\,W$ for the near-term system with $H = 30$, $K = 300$. With the clock frequency of 1 GHz, we would have $2.7 \times 10^{15}$ MAC operations per second, giving 10.26 W/PetaOPS (i.e. 10.26 fJ/MAC). For the long-term system with $H = 100$, $K = 1000$, and 1 GHz clock frequency, we would have $P \approx 206\,W$ and $100 \times 10^{15}$ MAC operations per second, giving 2.06 W/PetaOPS (i.e. 2.06 fJ/MAC). This is two order of magnitude more efficient compared to the Nvidia H100 GPU, which operates at approximately 100 W/PetaOPS[18].

The power consumption estimation presented in this section takes into account numerous realistic assumptions, including a 20 dB optical loss in the free-space optics setup. With better alignment of the optical setup, utilization of a wider spectral bandwidth, and future improvements to individual components, we believe that hyperspectral in-memory optical computing systems will eventually exceed Exa-OPS (i.e., $10^{18}$ MAC operations per second) while maintaining a power efficiency of 1 W/PetaOPS.

TABLE S2. **Estimated System Performance**

|  | Current (open-loop) | Near-term (closed-loop) | Long-term (closed-loop) |
|---:|:---:|:---:|:---:|
| Matrix Size | 128 × 64 | 300 × 300 | 1000 × 1000 |
| Hyperspectral Factor | × 1 (× 10) | × 30 | × 100 |
| Clock Frequency | 250 MHz[†] | 1 GHz | 1 GHz |
| Computational Throughput | 2.048 TOPS (20.48 TOPS) | 2.7 PetaOPS | 100 PetaOPS |
| Total Power Consumption | 11.9 W | 27.7 W | 206 W |
| Power Efficiency | 5.8 W/TOPS | 10.26 W/PetaOPS | 2.06 W/PetaOPS |

[†] We assume an external modulation and readout speed of 250 MHz.


\fi

\end{document}

\end{document}